\documentclass[twocolumn,showpacs,preprintnumbers,amsmath,amssymb]{revtex4}

\usepackage{amssymb}
\usepackage{amsmath}
\input{epsf}
\begin{document}

\title{Eulerian simulations of collisional effects on electrostatic plasma waves}

\author{Oreste Pezzi, Francesco Valentini, Denise Perrone and Pierluigi Veltri}
\affiliation{Dipartimento di Fisica and CNISM, Universit\`a della Calabria, 87036 Rende (CS), Italy}


\begin{abstract}
The problem of collisions in a plasma is a wide subject with a huge historical literature. In fact, the description of realistic plasmas is a tough problem to
attach, both from the theoretical and the numerical point of view, and which requires in general to approximate the original collisional Landau integral by
simplified differential operators in reduced dimensionality. In this paper, a Eulerian time-splitting algorithm for the study of the propagation of electrostatic
waves in collisional plasmas is presented. Collisions are modeled through one-dimensional operators of the Fokker-Planck type, both in linear and nonlinear form. The
accuracy of the numerical code is discussed by comparing the numerical results to the analytical predictions obtained in some limit cases when trying to evaluate the effects of collisions in the phenomenon of wave plasma echo and collisional dissipation of Bernstein-Greene-Kruskal waves. Particular attention is devoted to the study of the nonlinear Dougherty collisional operator, recently used to describe the collisional dissipation of electron plasma waves in a pure electron plasma column [M. W. Anderson and T. M. O’Neil, Phys. Plasmas {\bf 14}, 112110 (2007)]. A receipt to prevent the filamentation problem in Eulerian algorithms is provided by exploiting the property of velocity diffusion
operators to smooth out small velocity scales.
\end{abstract}

\date{\today}
\maketitle

\section{Introduction}
Unlike ordinary molecular gases, in realistic plasmas particle interactions are of infinite range, since the plasma dynamics is governed by the Coulomb force. The
problem of Coulomb collisions in a plasma has been the subject of a significant theoretical effort since the late thirties, when Landau \cite{landau36} proposed his
integro-differential operator for the description of particle collisions in a gas dominated by Coulombian interactions. The Landau operator is in general of the
Fokker-Planck type, i. e. it involves gradients of the particle distribution function in velocity space. In Ref. \cite{landau36} this operator was directly derived
from the Boltzmann collisional integral, by assuming the momentum exchange in each collision to be small. The form of the Landau operator can be also obtained from
Liouville equation, through the BBGKY hierarchy \cite{akhiezer86,swanson08}, by assuming the particle correlations to be ordered according to the small parameter $g$
(the plasma parameter) \cite{oneil65-1}.

In natural plasmas, such as the solar wind, the effects of collisions are usually considered negligible, since the interplanetary medium is a very rarefied gas (the
particle density is typically of the order of $10\ cm^{-3}$) and the mean free path of a particle traveling from the Sun to the Earth is estimated to be of the order
of $1AU$ (the Sun-Earth distance). Nevertheless, wave-particle interaction processes, at work in such a collisionless medium, can significantly shape the particle
distribution functions, creating sharp gradients and local deformations in velocity space. These distortions at small velocity scales can enhance locally the plasma
collisionality through the velocity gradients at play in the Landau collisional integral. A full understanding of the collisional phenomena in weakly collisional
plasma systems would surely represent a relevant step forward in the problem of heating of the interplanetary medium and of laboratory plasmas, so there is a strong need for numerical codes describing the effects of collisions.

The description of the plasma collisionality based on the complete Landau model is an extremely complicated problem to treat. From the analytical point of view, the
main difficulties come from the nonlinear nature of the Landau operator, while its multi-dimensionality limits any numerical approach to very low resolution
discretization in the velocity numerical domain. This is first due to the need of evaluating numerically the velocity derivatives of the distribution function in a
three-dimensional velocity space and secondarily to the fact that for each gridpoint of the phase space numerical domain a three-dimensional velocity integration
must be performed.

In 2000, Pareschi et al. \cite{pareschi00} proposed a spectral method for the numerical evaluation of the Landau collisional integral, based on the use of Fast
Fourier Transform (FFT) routines. This allows to obtain numerical solutions with $O(n\log_2{n})$ operations with respect to the usual $O(n^2)$ cost of a finite
difference scheme. However, in order to exploit this spectral approach, these authors periodized the distribution function in the velocity domain as well as the
Landau operator (for details, see Ref. \cite{pareschi00}), thus introducing unphysical effects of fake binary collisions. In 2002, Filbet and Pareschi
\cite{filbet02} proposed a time-splitting scheme for the numerical integration of the Landau-Poisson equations in 1D-2V  phase space configuration (1D in physical
space and 2D in velocity space), where the spectral method introduced in Ref. \cite{pareschi00} is used to evaluate the Landau integral. Two years later, Crouseilles
and Filbet \cite{crouseilles04} extended the above mentioned 1D-2V Landau-Poisson code to the 1D-3V geometry and analyzed numerically the competition between Landau
damping and collisional dissipation in the propagation of electrostatic plasma waves. Despite the use of FFT routines, the numerical simulations in Refs.
\cite{filbet02,crouseilles04} have been run with a quite limited resolution in phase space (typically, 50 gridpoints in the physical domain and 30-60 gridpoints in
each velocity direction).
\begin{figure}
\epsfxsize=8cm \centerline{\epsffile{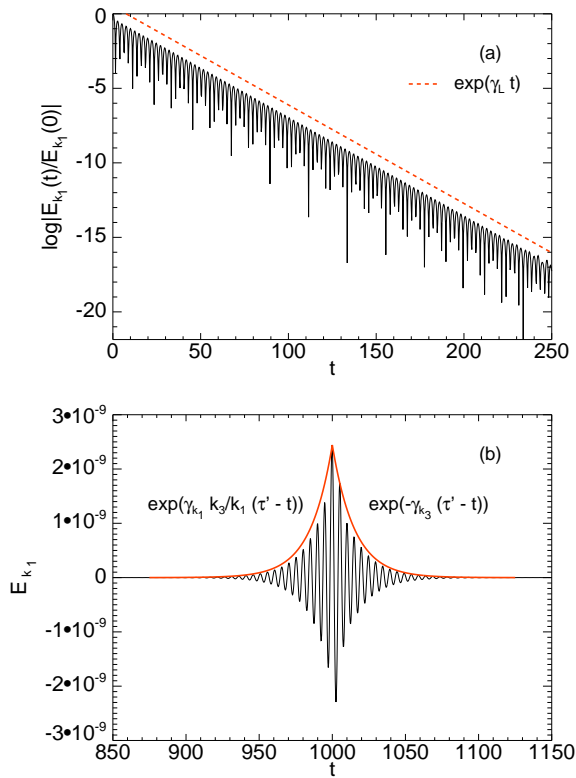}}   % FIGURE N.1
\caption{(Color online) (a) Time evolution of the logarithm of the electric field Fourier component $|E_{k_1}|$ normalized to its initial value. The red-dashed line
represents the theoretical expectation for Landau damping. (b)  $E_{k_1}$ as a function of time, in the range $t=[850,1150]$. The red-solid curves represent the
theoretical predictions for rise and fall of the wave echo.}
\label{fig1}
\end{figure}

In this paper, we present a 1D-1V Eulerian collisional Vlasov-Poisson code, in which the effects of collisions are included in the right-hand side of the Vlasov
equation through one-dimensional differential operators (both linear and nonlinear). In this algorithm, the time advance scheme for the particle distribution function is realized along the lines of the time-splitting method discussed in Ref. \cite{filbet02}. The reduced phase space dimensionality together with the use of simplified 1D
differential collisional operators allow us to perform high resolution numerical simulations which can have important applications as a theoretical support to
experiments on electrostatic plasma waves, such as those performed in Penning-Malmberg machines with nonneutral plasmas.

We considered three differential collisional operators of the Fokker-Planck type (whose details will be discussed in next section). In order to demonstrate the
accuracy of our numerical approach, we compared the numerical results with the analytical predictions for the effects of collisionality on the phenomenon of wave
plasma echoes \cite{gould67} and on the propagation of Bernstein-Greene-Kruskal (BGK) waves \cite{bgk}, in some limit cases. We focused in particular on the numerically description of the properties of the Dougherty collisional operator \cite{dough}, an intrinsically nonlinear differential operator which has been recently considered
to study the effects of collisions in the propagation of electron plasma waves in nonneutral plasmas \cite{anderson07}. Finally, by exploiting the properties of the
differential collisional operators of dissipating the small velocity scales through diffusive processes, we propose a receipt to prevent collisional Eulerian
simulations to suffer from the so-called filamentation problem.

This paper is organized as follows. In Section II, we discuss the mathematical model adopted to describe particle collisions and the numerical approach implemented.
In Section III, we test the accuracy of our numerical algorithm by comparing the numerical results to the analytical predictions for the collisional dissipation and
deformation of plasma wave echoes and for the collisional damping of BGK structures. In Section IV, we present the numerical results of the collisional Eulerian code
when the Dougherty operator is implemented. Section V is focused on the filamentation problem of Eulerian algorithms. We finally conclude and summarize in Section
VI.

\begin{figure}
\epsfxsize=8cm \centerline{\epsffile{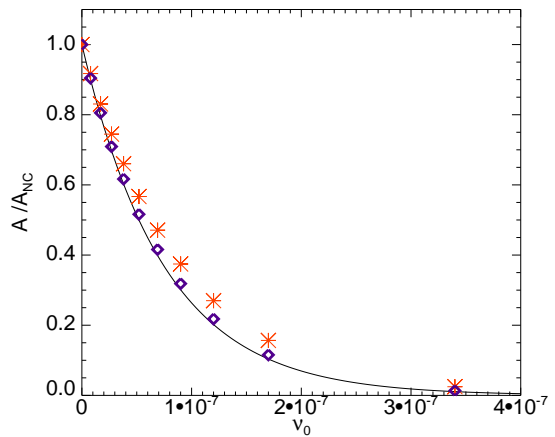}}   % FIGURE N.2
\caption{(Color online) Echo amplitude $A$ (normalized to $A_{_{NC}}$) as a function of $\nu_0$. The red stars indicate numerical results obtained through the EK1
time scheme, while purple diamonds represent simulations performed by making use of the RK4 time scheme. The black solid line is the theoretical prediction in Eq.
(\ref{d2amp}).}
\label{fig2}
\end{figure}
\section{Mathematical model and numerical approach}
In the framework of kinetic plasma theory, the propagation of electrostatic plasma waves in the absence of particle collisions can be described by the
Vlasov-Poisson equations, in the simplified 1D-1V phase space geometry. In the present paper, we analyze in detail the properties of different one-dimensional
collisional operators and their effects on the propagation of plasma waves, by means of Eulerian kinetic simulations. This has been done by including in the
right-hand side of the Vlasov equation different collisional operators (both linear and nonlinear) of the Fokker-Planck type. In our analysis only electron-electron
collisions are taken into account, while the effects of electron-ion interactions and ion-ion interactions have been neglected.

The basic equations considered in our treatment can be written in the following dimensionless form:
\begin{eqnarray}
& & \frac{\partial f}{\partial t}+v\frac{\partial f}{\partial x}+\frac{\partial\phi}{\partial x}\frac{\partial f}{\partial v}=C(f,f)\label{vlaseq}\\
&-&\frac{\partial^2 \phi}{\partial x^2} =  1 -\int f dv \label{poiseq}
\end{eqnarray}
where $f$ is the electron distribution function, $\phi$ is the electrostatic potential and $C(f,f)$ is a generic collisional operator, which is, in general, an
integro-differential functional of $f$. Due to their inertia, the ions are considered as a motionless neutralizing background of constant density $n_0=1$. In the
previous equations, time is scaled to the inverse electron plasma frequency $\omega_{pe}$, velocities to the initial electron thermal speed $v_{th,e}$; consequently, lengths
are normalized by the electron Debye length $\lambda_{De}=v_{th,e}/\omega_{pe}$. For the sake of simplicity, from now on, all quantities will be scaled using the
characteristic parameters listed above.

We consider in detail three different 1D collisional operators. The first one is the linear Zakharov-Karpman (ZK) operator \cite{ZK63}, whose form is essentially equivalent to that discussed by Lenard and Bernstein in 1958 \cite{lenard58}. The ZK operator has been obtained by linearizing the original Landau integral in the resonant region (that is for velocities close to the phase velocity $v_\phi$ of the wave) and assuming distribution functions close to Maxwellians. The ZK operator has been considered in
Ref.\cite{ZK63} to describe the collisional damping of BGK waves \cite{bgk}, and its expression, in dimensionless units, can be written as follows:
\begin{equation}
 C(f) =  \nu\frac{\partial}{\partial v}\left( \frac{\partial f}{\partial v} + v f \right)
 \label{ZKeq}
\end{equation}
where $\nu=\nu_0[6(v_{th,e}/v_\phi)^3]$ is the collision frequency, $\nu_0$ is related to the plasma parameter $g$ (the number of particles in a Debye sphere),
according to the equation $\nu_0=g\ln{\Lambda}/8\pi$; $\ln\Lambda$ is the Coulombian logarithm, related to $g$ as $\ln\Lambda\simeq -(\ln{g})/3$.

The second operator under consideration has been analyzed by O'Neil in 1968 \cite{oneil68} in studying the effects of Coulomb collisions on plasma wave echoes. For this reason, we will refer to this operator as the O'Neil (ON) operator. The form of the ON operator has been derived from the Landau integral through a linearization procedure, by assuming distribution functions not far from the Maxwellian shape; the first velocity derivative of the distribution function, present in the Landau operator, has been neglected by O'Neil assuming that in the large time limit the term containing the second velocity derivative of $f$ is dominant. Moreover, in the ON operator, the collision frequency has been obtained without specializing for resonant electrons, as in the ZK case. The explicit expression of the ON operator is the following:
\begin{equation}
 C(f)= \frac{\partial^2}{\partial v^2} \left( \nu f \right)
 \label{ONeq}
\end{equation}
where $\nu$ can be in general a function of $v$ and it has the velocity dependence appropriate to Coulomb collisions \cite{oneil68}:
\begin{equation}
 \nu(v) = - \nu_0\frac{1}{v} \frac{\partial}{\partial v}
 \left[\frac{1}{v} erf\left(\frac{v}{\sqrt{2}}\right) \right] \ 
 \label{d2on}
\end{equation}
\begin{figure}
\epsfxsize=8cm \centerline{\epsffile{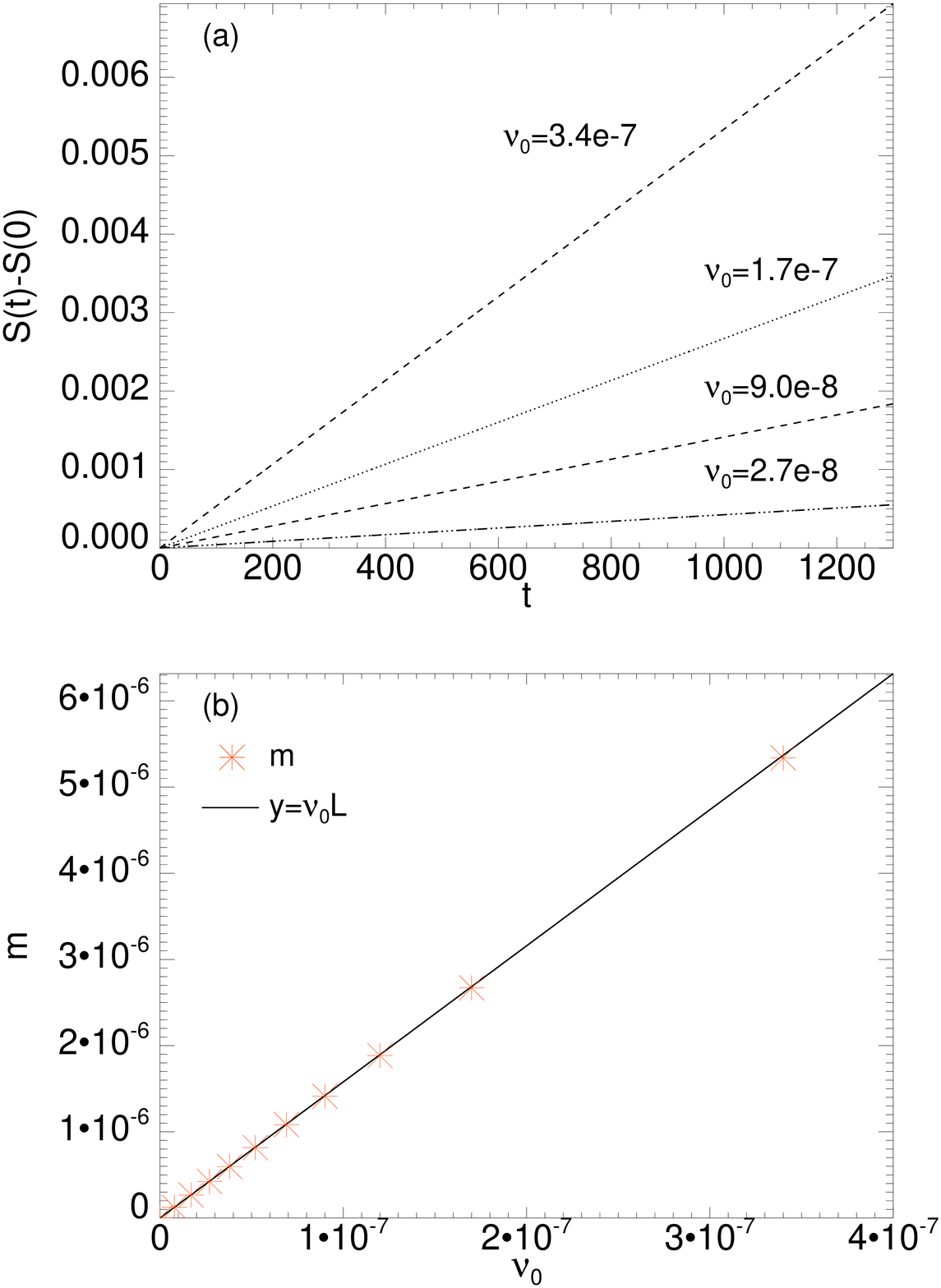}}   % FIGURE N.3
\caption{(Color online) (a) Time evolution of entropy variation $\Delta S$ for four simulations corresponding to different values of $\nu_0$ ($\nu_0=3.4\times
10^{-7}, 1.7 \times 10^{-7}, 9.0\times 10^{-8}, 2.7\times 10^{-8}$). (b) The slope $m$ of the linear time growth of $\Delta S$ as a function of $\nu_0$ (red stars);
the black-solid line indicates the curve $y=\nu_0 L$.}
\label{fig3}
\end{figure}

Finally, we will analyze the properties of the Dougherty (DG) operator \cite{dough}, first derived in 1967. The DG operator is in the typical nonlinear Fokker-Planck form and it has been derived by requiring the conservation of mass, energy and momentum and that a generic Maxwellian is the unique equilibrium solution. The expression of the DG operator is written as:
\begin{equation}
 C(f,f) = \nu(n,T) \frac{\partial}{\partial v} \left[ T(f) \frac{\partial f}{\partial v} +
\left( v - U(f) \right) f \right]
 \label{doueq}
\end{equation}
where $n=\int f dv$ is the electron density, $U=(\int vf dv)/n$ is the electron mean velocity, $T=(\int (v-U)^2 f dv)/n$ the electron temperature and the collision
frequency $\nu$ has the form:
\begin{equation}
 \nu(n,T) = \nu_0\frac{n}{T^{3/2}}
 \label{dougnu}
\end{equation}
The DG operator is explicitly nonlinear and it has been recently used to study the collisional damping of Trivelpiece-Gould waves in non-neutral plasmas confined in
a Penning-Malmberg apparatus \cite{anderson07}.

We solve numerically  Eqs. (\ref{vlaseq})-({\ref{poiseq}) through a Eulerian code based on finite difference
methods for the evaluation of spatial and velocity derivatives of $f$ on the gridpoints.

Time evolution  of the distribution function is approximated by using a splitting scheme appropriately designed by Filbet et al. \cite{filbet02} for collisional
Eulerian codes that decomposes the evolution of $f$ in three different steps. We summarize this splitting scheme, by explicitly describing a single time step
($\Delta t$) :
\begin{enumerate}
 \item $\Delta t/2$ transport step $\rightarrow \partial_t f+ v \partial_x f + \partial_x \phi\; \partial_v f= 0$
 \item $\Delta t$ collision step $\rightarrow \partial_t f = C(f,f)$
 \item $\Delta t/2 $ transport step $\rightarrow \partial_t f+ v \partial_x f + \partial_x \phi; \partial_v f= 0$
\end{enumerate}
The collision step has been performed using two different explicit time schemes (coupled to finite difference schemes for the derivatives of $f$), whose performances
and accuracy will be discussed and compared in the next section:

\begin{enumerate}
 \item Eulerian first order scheme in time coupled to a second order centered finite difference scheme for $f$ derivatives, to which we will refer as EU1 scheme;
 \item Runge-Kutta fourth order scheme in time coupled to a sixth order centered finite difference scheme for $f$ derivatives, to which we will refer as RK4.
\end{enumerate}

Each transport step is in turn composed by three sub-steps, a first half-step advection in physical space followed by a full-step advection in velocity space and
then by an additional half-step advection in physical space, according to the time splitting scheme first proposed by Cheng and Knorr in 1976 \cite{cheng76} (see
also Refs. \cite{mangeney00,valentini05,valentini07}). The Poisson equation for the electrostatic potential is solved after the first spatial advection step. Being
$\Delta t'=\Delta t/2$ the time step for the transport advance, a single transport step can be summarized as follows:
\begin{enumerate}
 \item $\Delta t'/2$ $x$-advection $\rightarrow \partial_t f+ v \partial_x f=0 \rightarrow$ Poisson
 \item $\Delta t'$ $v$-advection $\rightarrow \partial_t f+ \partial_x\phi\; \partial_v f=0$
 \item $\Delta t'/2$ $x$-advection $\rightarrow \partial_t f+ v \partial_x f=0$
\end{enumerate}
\begin{figure}
\epsfxsize=8cm \centerline{\epsffile{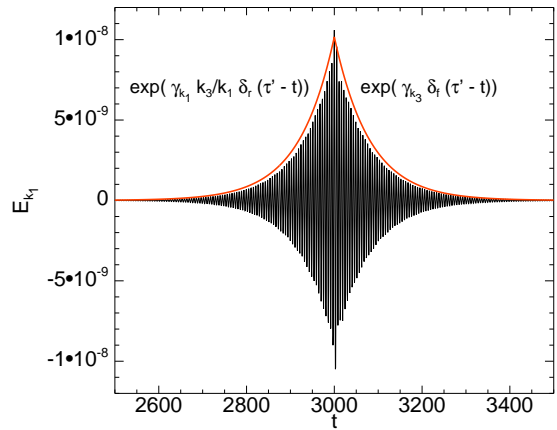}}   % FIGURE N.4
\caption{(Color online) The spectral electric field $E_{k_1}$ as a function of time, in the range $t=[2500,3500]$ (black line). Red lines represent the theoretical
expectations for echo raise and fall coefficients from O'Neil theory in Eqs. (\ref{ron})-(\ref{fon}).}
\label{fig4}
\end{figure}

Both $x$-advection and $v$-advection have been performed numerically through an upwind finite difference scheme correct up to third order in spatial and velocity
mesh size (this is the so-called Van Leer scheme \cite{mangeney00}, which has been successfully employed in the collisionless case in many research works, as, for example, in Refs. \cite{valentini11, valentini12, valentini13, perrone13}). Since periodic boundary conditions have been imposed in the spatial numerical domain, we solve Poisson
equation through a standard FFT routine. In the numerical velocity domain, $f$ is set equal to zero for $|v|>v_{max}$, where $v_{max}=6 v_{th,e}$. Typically, the
numerical phase space domain is discretized with $N_x=128$ gridpoints in the physical domain $D_x=[0,L]$ and $N_v=1301$ gridpoints in velocity domain
$D_v=[-v_{max},v_{max}]$. The time step $\Delta t$ is chosen in such a way to satisfy Courant-Friedrichs-Levy condition \cite{peyret83}, for the numerical stability
of time explicit finite difference algorithms.

\section{Accuracy tests on the numerical collisional algorithm}
In this section, we systematically compare the numerical results of our collisional Eulerian code with the analytical predictions concerning the effects of Coulomb
collisions on electrostatic plasma waves. We consider in detail two different physical phenomena: the first is the dissipation and deformation of plasma wave echoes
\cite{gould67,malmberg68} produced by Coulomb collisions, analytically treated by O'Neil in 1968 \cite{oneil68}; the second concerns the collisional damping of BGK
wave structures discussed by Zakharov and Karpman in 1963 \cite{ZK63} and numerically revisited in 2008 \cite{valentini08} through collisional Particle in Cell
simulations. These physical phenomena have been reproduced by making use of the collisional Eulerian code, thus showing that the time splitting method described in
detail in the previous section produces results that well agree with analytical expectations.

\subsection{Effect of collisions on plasma wave echoes}\label{echo}
In the absence of collisions, a small amplitude electrostatic perturbation imposed on a plasma initially at equilibrium is dissipated in time by Landau damping
\cite{landau46}. When an electric perturbation of spatial dependence $e^{-ik_1x}$ ($k_1$ being the wavenumber) is excited in the plasma and then is damped away, it
leaves a modulation in the distribution function in the form $f_1(v)\exp{[-ik_1x+ik_1vt]}$, this being usually called the ballistic term. While no electric field is
associated with the ballistic term in the large time limit (since the velocity integral of $f_1(v)$ gets null by phase mixing), the exponential
$\exp{[-ik_1x+ik_1vt]}$ in the perturbed distribution function never dies, getting more and more oscillatory in time. In the same way, if a second wave of spatial
dependence $e^{ik_2x}$ is launched in the plasma at time $\tau$, it will be Landau damped in time and will leave a first-order modulation in the distribution
function of the form $f_2(v)\exp{[-ik_2x+ik_12vt]}$ plus a second-order contribution that can be written as $f_1(v)f_2(v)\exp{[i(k_2-k_1)x+ik_2v\tau-i(k_2-k_1)vt]}$.
It is clear from previous expression that at $t=\tau [k_2/(k_2-k_1)]$ the coefficient of $v$ in the exponential will vanish; as a consequence, the velocity integral
of the second-order perturbed distribution function does not phase mix to zero in this case, thus producing the appearance of a third electric signal with wavenumber
$k_3=k_2-k_1$ in the plasma, called echo \cite{gould67}. It is worth noting that the echo is a second order effect in the perturbation amplitudes.
\begin{figure}
\epsfxsize=9cm \centerline{\epsffile{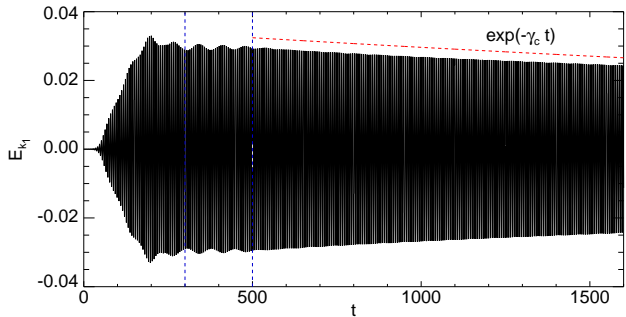}}   % FIGURE N.5
\caption{(Color online) Time evolution of $E_{k_1}$. The left blue-dashed line represents the time instant at which the driver is turned off, while the right
blue-dashed line indicates the time instant at which collisions are turned on. Finally, the red-dashed line indicates the theoretical expectation for collisional
damping, as obtained by Zakharov and Karpman in Ref. \cite{ZK63}.}.
\label{fig5}
\end{figure}

The physics of plasma wave echoes has been extensively investigated from the numerical point of view, by means of collisionless Eulerian codes
\cite{nocera99,galeotti06} and represent an excellent benchmark to test the accuracy of numerical algorithms. We first used our Eulerian code in the collisionless
approximation to reproduce the generation of the plasma wave echo. For this simulation in linear regime, at t=0 the electrons have homogeneous density $n_0$ and
Maxwellian distribution of velocities. We impose on this equilibrium a sinusoidal density perturbation with wavenumber $k_1=0.4$ ($k_1$ is the fundamental
wavenumber, i. e. $\lambda_1=2\pi/k_1$ is the maximum wavelength that fits in the simulation box) and amplitude $n_1=4\times 10^{-5}$ (corresponding to an electric
field amplitude $E_1=10^{-4}$). In Figure \ref{fig1} (a), we report the time evolution of the logarithm of the electric field Fourier component $E_{k_1}$ (normalized
to its initial amplitude) up to the time $t=250$. The red-dashed line in this figure represents the theoretical prediction for the wave damping rate obtained through
a numerical code that solves for the roots of the electrostatic dielectric function, by evaluating numerically the complex Landau integral \cite{krall86}; one can
notice a very good agreement between numerical and theoretical results.

In the same simulation, at time $\tau=500$ a second electric perturbation is launched with wavenumber $k_2=0.8$ and amplitude $E_2=E_1$. According to the
predictions in Ref. \cite{gould67}, a plasma wave echo with wavenumber $k_3=k_2-k_1=0.4=k_1$ appears at $t=\tau'=\tau [k_2/(k_2-k_1)]=1000$. In Figure \ref{fig1}
(b), we show the time evolution of the electric field Fourier component $E_{k_1}=E_{k_3}$ in the time interval $850\leq t \leq 1150$: here, a small amplitude
electric signal (the echo) grows in time, reaches a peak value at $t=\tau'=1000$ and then is Landau damped away. The red-solid curves in this figure represent the
analytical predictions for growth and damping rates of the echo as derived in Ref. \cite{gould67}, where $\gamma_{k_1}$ and $\gamma_{k_3}$ are the absolute value of
the Landau damping coefficient for wavenumber $k_1$ and $k_3$ respectively. We point out that for this simulation, since  $k_1=k_3$ (and consequently
$\gamma_{k_1}$=$\gamma_{k_3}$), the echo results symmetric around $t=\tau'=1000$. The analytical predictions for the collisionless raise and fall coefficients of the
echo give a value $\Gamma_r^{(th)}=|\Gamma_f^{(th)}|\simeq 0.06613$; the numerical simulation provides $\Gamma_r^{(num)}=|\Gamma_f^{(num)}|\simeq 0.06632$, with a
relative error of about $0.3\%$.

Finally, we point out that for this simulation, the conservation of the Vlasov invariants is highly satisfactorily: total energy variations are of the order of
$\simeq 6\times 10^{-7}\%$, for mass variations we get $\simeq 10^{-13}\%$ and for entropy variations $\simeq 7.15\times 10^{-10}\%$.

As a next step, we repeated the same simulation described above turning on the collisional effects described by the ON operator. In his 1968 paper \cite{oneil68},
O'Neil studied the effects of Coulomb collisions, first assuming $\nu = \mathit{const}=\nu_0$ and then when $\nu$ is a function of $v$. In the first case, he showed
that the shape of the echo remains unchanged, while its amplitude decreases exponentially as the value of $\nu_0$ increases, according to the following equation:

\begin{equation}
  \frac{A}{A_{_{NC}}} = \exp \left(\frac{-\nu_0 k_{1}^2 k_{2} \tau^3}{3 k_{3}}\right)
 \label{d2amp}
\end{equation}
where $A$ is the echo amplitude in the presence of collisions and $A_{_{NC}}$ is the echo amplitude in the collisionless case. This result has been obtained in the
limit where the time between the launching of the first electric pulse and the appearance of the echo is much larger than the time for Landau damping (i. e.,
$\eta=1/(\gamma_k \tau') \ll 1$) and under the approximation of having negligible collisional damping while electric fields are present ($\zeta=\nu_{eff}/\gamma_k\ll
1$, where $\nu_{eff}$ can be estimated as $\nu_0 k^2 t^{*2}$ \cite{oneil68}, $t^*$ being the time between the appearance of the echo and the time at which each
electric pulse is launched). For the parameters of our simulation, we get $\eta\simeq 10^{-2}$, while $\zeta\simeq 0.3$ for the first pulse ($k=k_1=0.4$) and
$\zeta\simeq 0.22$ for the second pulse ($k=k_2=0.8$); we point out that the above values of $\zeta$ have been calculated for the less favorable situation, that is
for the maximum value of $\nu_0$ considered in our numerical study. Therefore, we can conclude that our numerical analysis has been performed within the limits of
validity of the analytical results in Ref. \cite{oneil68}.
\begin{figure}
\epsfxsize=8cm \centerline{\epsffile{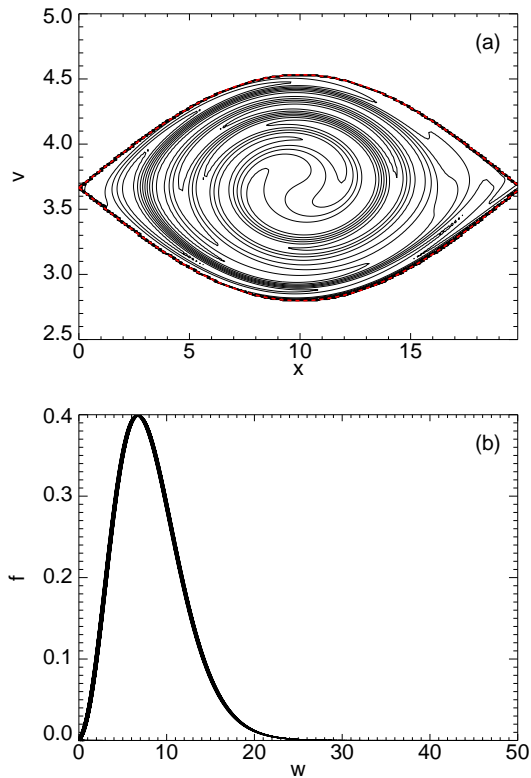}}   % FIGURE N.6
\caption{(Color online) (a) Contour plot of the distribution function of trapped electrons at $t=500$. The red-dashed curves represent the analytical separatrices
between regions of trapped and free motion. (b) Scatter plot of the electron distribution function at $t=500$ as a function of energy in the wave frame.}
\label{fig6}
\end{figure}

We performed 11 simulations with different values of $\nu_0$ in the range $7.9\times 10^{-9}\le \nu_0\le 3.4\times 10^{-7}$. In Fig. \ref{fig2} we plot the ratio
$A/A_{_{NC}}$ as function of $\nu_0$. The black solid line represents the analytical prediction in Eq. (\ref{d2amp}); the red stars represent the numerical results
when the EU1 scheme is used to perform the collisional step (see previous section), while the purple diamonds have been obtained from the numerical simulations when
the RK4 scheme has been employed. It can be easily seen in this figure that a very good agreement between theory and simulation is obtained when the RK4 scheme is
used, while numerical and analytical curves depart a little when the less precise EU1 scheme is adopted. For these reasons, all numerical experiments discussed in
the following have been performed employing the more accurate RK4 scheme for the collisional step.

When $\nu$ does not depend on $v$ ($\nu=\nu_0$) and the amplitude of the electric perturbations is small, a simple calculation, performed assuming $f$ to be close
to a homogeneous Maxwellian, shows that the entropy of the system, defined as $S(t)=-\int f\ln{f} dx\; dv$, increases linearly in time following the equation $\Delta
S=S(t)-S(0)=\nu_0 L t$. According to this analytical expectation, from the results of our numerical experiments we recovered a linear increase of the entropy in
time. The top panel of Fig. \ref{fig3} shows the time variation of the entropy for four simulations corresponding to four different values of $\nu_0$. As it is clear
from this figure, the numerical entropy variation depends on time as $\Delta S=m t$, where $m=m(\nu_0)$. In order to compare the numerical results to the analytical
expectations, we evaluated $m$ for each of the 11 simulations and reported the results for $m$ versus $\nu_0$ as red stars in the bottom plot in Fig. \ref{fig3}. In
the same plot, the black-solid line indicates the analytical expectation for $m$; a very good agreement between numerical and theoretical results is clearly visible.

As discussed in Ref. \cite{oneil68}, when $\nu$ depends on $v$ as in Eq. (\ref{d2on}), the collisional damping can modify the shape of the echo. In particular,
O'Neil showed that, under the approximation of large velocities with respect to the electron thermal speed ($\omega/k\gg v_{th,e}$, $\omega$ being the real part of
the wave frequency), collisional damping reduces the raise of the echo by the factor:
\begin{equation}
 \delta_{r} = \exp \left[ - \nu_{0}\left(\frac{\omega_1}{k_1}\right) \frac{k_1^2k_2\tau^3}{3k_3}
 \right]
 \label{ron}
\end{equation}
while the fall of the echo by the factor:
\begin{equation}
 \delta_{f} = \exp \left[ - \nu_{0}\left(\frac{\omega_3}{k_3}\right) \frac{k_1^2k_2\tau^3}{3k_3}
\right]
\label{fon}
\end{equation}
where $\omega_1$ and $\omega_3$ are the wave frequencies corresponding to wavenumbers $k_1$ and $k_2$ respectively. In the specific case in which
$k_2=2k_1\Rightarrow k_3=k_1$, one gets $\delta_r = \delta_f$ and the shape of the echo remains unchanged.
\begin{figure}
\epsfxsize=9cm \centerline{\epsffile{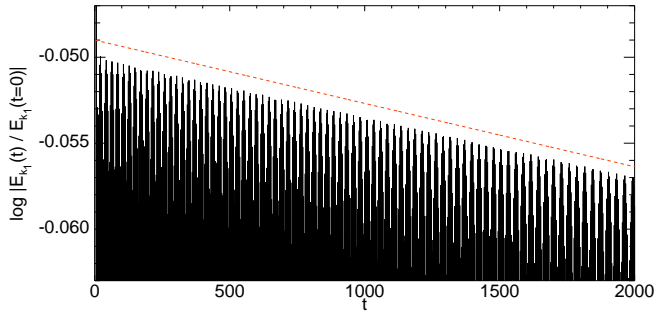}}   % FIGURE N.7
\caption{(Color online) Time evolution of the logarithm of the electric field Fourier component $|E_{k_1}|$ normalized to its initial value. The red-dashed line
represents the theoretical expectation for wave damping in the second of Eqs. (\ref{infplasma}).}
\label{fig7}
\end{figure}

In order to test our Eulerian collisional code, we repeated the wave echo simulation using a slightly different set of parameters with respect to the previous
simulation. The values of the wavenumbers $k_1$ and $k_2$ has been set $k_1=0.3$ and $k_2=0.6$ (consequently $k_3=k_1=0.3$). The time at which the second electric
pulse is launched is $\tau = 1500$, consequently the time of the echo is $\tau'=3000$. We used Eq. (\ref{d2on}) for the velocity dependence of $\nu$ following O'Neil
\cite{oneil68}. Figure \ref{fig4} shows the time evolution of the electric field spectral component $E_{k_1}=E_{k_3}$ in the time interval $2500 \le t \le 3500$. The
appearance of the echo is clearly visible with a peak value at $t=\tau'$. The red-solid lines in this plot represent the theoretical prediction for growth and
damping of the echo. A very good agreement is visible from this plot; to be more quantitative, the relative error between numerical and theoretical results is of
about $1.5\%$. This slight discrepancy can be due to the fact that for the values of the wavenumbers chosen in this numerical experiment we get $\omega_1/k_1\simeq
3.87 v_{th,e}$ and $\omega_2/k_2\simeq 2.6 v_{th,e}$; therefore the approximation of having large velocities with respect to $v_{th,e}$, under which the theoretical
predictions in Eqs. (\ref{ron})-(\ref{fon}) have been obtained, is only weakly satisfied in our numerical analysis.

\subsection{Collisional damping of BGK modes}\label{ZKBGK}
In this section we present a numerical study of the effects of collisions, modeled according to the ZK operator discussed previously, on the stability of BGK modes. The BGK modes \cite{bgk} are nonlinear stationary solutions of the collisionless Vlasov-Poisson equations characterized by a population of trapped particles and whose nonlinearity
manifests as a distortion of the particle distribution function in the region around the wave phase speed. When collisions are neglected, the main manifestation of this distortion is the formation of a plateau (region of zero slope) in the particle velocity distribution in the vicinity of the phase velocity of the wave; because of the presence of this plateau, the electric oscillations result completely undamped in time. The effect of collisions, which tend to restore the thermodynamic equilibrium, is to smooth out this plateau, thus destroying the region of trapped particles and driving the velocity distribution towards a Maxwellian shape. As a consequence, the electrostatic oscillation will undergo collisional damping.

This effect has been studied analytically by Zakharov and Karpman in 1963 \cite{ZK63}, which modeled particle collisions through the one-dimensional linear velocity
diffusion operator in Eq. (\ref{ZKeq}). These authors assumed the wave damping due to collisions to be small and derived a simple equation for the dissipation of the
electric perturbation amplitude in the form $E(t)\simeq E_0e^{-\gamma_c t}$. The damping coefficient $\gamma_c$ has the following (dimensionless) expression (see
also Ref. \cite{valentini08}):

\begin{equation}
 \gamma_c = \frac{2}{3} \left[\frac{12\left(7\pi+6\right)}{16\pi}\right] v_{\phi}^4 e^{-v_{\phi}^2/2} \nu \left(\frac{E_0}{k}\right)^{-\frac{3}{2}}
 \label{gceq}
\end{equation}
in which $v_\phi=\omega/k$ is the wave phase velocity and $E_0$ is the initial perturbation amplitude.
\begin{figure}
\epsfxsize=8cm \centerline{\epsffile{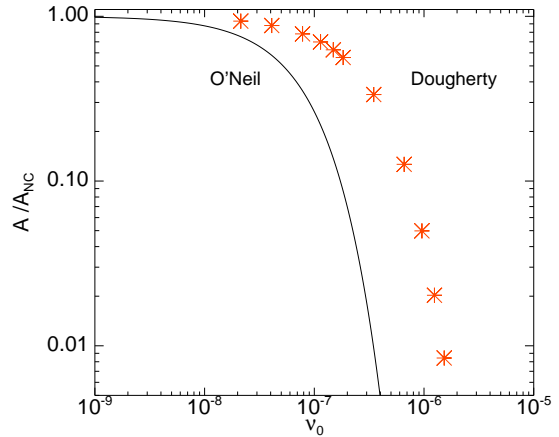}}   % FIGURE N.8
\caption{(Color online) Log-log plot of the echo amplitude $A$ (normalized to $A_{_{NC}}$) as a function of $\nu_0$. The red stars indicate simulations performed
with the DG operator, while the black-solid line is the theoretical curve by O'Neil in Eq. (\ref{d2amp}).}
\label{fig8}
\end{figure}

In order to test the accuracy of our collisional Eulerian algorithm we have compared the numerical results for the collisional damping of BGK modes with the
analytical prediction in Eq. (\ref{gceq}). Our numerical experiment is divided into two steps:
\begin{enumerate}
 \item The first step consists in the excitation of the BGK wave through an external forcing electric field in the absence of collisions: we drive an initially
homogeneous and Maxwellian plasma of electrons (and fixed ions) through a sinusoidal small amplitude external electric field that traps resonant electrons and
flattens the velocity distribution in the vicinity of $v_\phi$. This turns off Landau damping, and finally leads to the excitation of the BGK oscillations.
 \item Once the BGK structure has been created, the driver is turned off smoothly and the electric field rings at a nearly constant amplitude in time. Then, as a second step, we
turn on collisions and observe the collisional damping of the electric field amplitude of the BGK wave.
\end{enumerate}

The external driver electric field is taken to be of the form:
\begin{equation}
 E_D (x,t) = E_D^{max}\left[ 1 + \left( \frac{t-\tau}{\Delta\tau}\right)^p\right]^{-1}\sin{(k_1 x - \omega_D t)}
 \label{drivBGK}
\end{equation}
where $\tau=150$, $\Delta\tau=100$, $p=16$. $k_1=\pi/10$ is the fundamental wavenumber and $\omega_D$ the frequency of the driver, corresponding to the Langmuir
frequency $\omega_L=(1+3k_1^2)^{1/2}$. The driver is turned on and off adiabatically. The driver amplitude is near $E_D^{max}=2.4\times 10^{-3}$ for about a trapping
period $\tau_{_T}=2\pi/\sqrt{k_1E_D^{max}}$ \cite{oneil65} and is near zero again by $t_{off}=300$. Collisions are turned on for $t>500$.

Figure \ref{fig5} shows the time evolution of the spectral electric field $E_{k_1}$. The first vertical blue-dashed line in this plot indicates the time at which
the driver is turned off ($t\simeq 300$), while the second blue-dashed line corresponds to the time at which collisions are turned on ($t = 500$). By looking first
at the collisionless part of the simulation, one notices that during the driving process the electric field grows to large amplitude, then oscillates around a
constant value $E_{_{BGK}}\simeq 0.06$ ($E_{_{BGK}}=2E_{k_1}^{sat}$, $E_{k_1}^{sat}$ being the saturation amplitude of $E_{k_1}$ in the time interval $300\le t\le
500$) and rings at this amplitude for $300\le t \le 500$.

In order to show that the one created by the external driver actually is a BGK structure, in Fig. \ref{fig6} (a) we reported the $x-v$ contour lines of the
distribution function of trapped electrons at $t=500$, that is right before collisions are turned on. This trapped electron distribution has been obtained by
selecting velocities in the interval $v^-(x)\leq v \leq v^+(x)$, where $v^{\pm}(x)= v_\phi\pm\sqrt{2[w(x,t=500)+\phi(x,t=500)]}$ are the separatrices between free
and trapped regions, while $w=(v-v_\phi)^2/2-\phi(x,t=500)$ is the particle energy in the wave frame at $t=500$. Here, a vortical structure, typical signature of the
presence of a trapped particle population, can be clearly recognized. The red-dashed curves in this plot indicate the $v^{\pm}(x)$ curves. At the bottom in the same
figure [panel (b)], we plotted the distribution function $f$ at $t=500$ as a function of $w(x,t=500)$. In other words, for each $(x,v)$ in the numerical domain we
plotted $f(x,v)$ as a function of $w(x,v)$, resulting in the single curve shown in Fig. \ref{fig6} (b). The points in this scatter plot clearly lie on a single curve
and this shows that $f$ depends on the energy $w$ alone, as expected for a BGK distribution. Moreover, using the electrostatic potential from the simulation
$\phi(x,t=500)$ in the BGK formalism, for the value of the wave phase speed obtained by the Fourier analysis of the numerical signal in the time interval $300 \leq t
\leq 500$ ($v_\phi\simeq 3.6652$), one can get the self-consistent BGK solution for the distribution function. This has been compared to the numerical distribution
function at $t=500$, giving a relative error of $\simeq 0.5\%$. Therefore, we can conclude that a BGK structure has been generated in the simulation by the external
driver electric field.
\begin{figure}
\epsfxsize=9cm \centerline{\epsffile{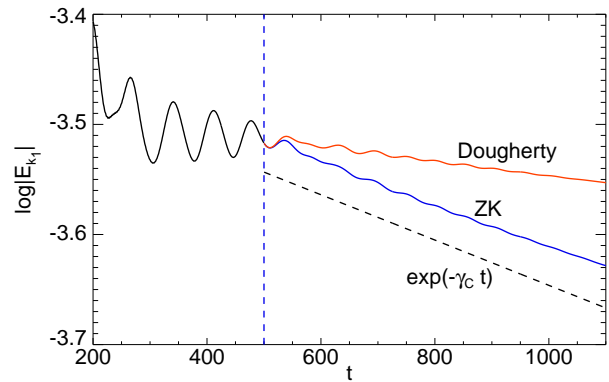}}   % FIGURE N.9
\caption{(Color online) Logarithm of $|E_{k_1}|$ versus time. The vertical blue-dashed line represents the time instant at which collisions are turned on. The blue
and red solid curves represent the results of two different simulations in which collisions are modeled through the ZK and the DG operators, respectively. The
black-dashed line represents the expectation for collisional damping in Eq. (\ref{gceq}).}
\label{fig9}
\end{figure}

Once the BGK structure has been created and the external driver has been turned off, at $t=500$ the effects of collisions come into play in the simulation. The
collision frequency $\nu$ in Eq. (\ref{ZKeq}) has been assigned the small value $\nu=2 \times 10^{-5}$. As it can be easily seen in panel (a) of Fig. \ref{fig5}, for
$t>500$ the wave amplitude slowly decays in time. The red-dashed line in this plot indicates the analytical expectation for collisional wave damping, with
$\gamma_c$ evaluated from Eq. (\ref{ZKeq}) using the numerical value of the wave phase speed $v_\phi\simeq 3.6652$, $k_1$ as the wavenumber and $E_{_{BGK}}$ as $E_0$
(the initial perturbation amplitude). This figure again shows a very good agreement between numerical results and analytical predictions.

\section{Numerical study of collisional effects modeled through the Dougherty operator}
In this section we describe the numerical results of the Eulerian kinetic code when the effects of collisions are modeled through the nonlinear DG operator in Eq.
(\ref{doueq}). We first discuss the numerical results that reproduce the collisional damping of plasma waves in linear regime. In 2007, Anderson and O'Neil
\cite{anderson07} derived the expression for the complex frequency of electron plasma waves (the so-called Trivelpiece-Gould waves \cite{trivelpiece59}) in a pure
electron plasma column confined in a Penning-Malmberg apparatus, when collisional effects are taken into account through the DG operator. For large wave phase
velocities, where Landau damping is negligible, the authors showed that the complex wave frequency $\omega$ has the following (dimensionless) expression:
\begin{equation}
 \omega=\frac{k_z}{k}\left[1+\frac{3}{2}k^2\left(1+\frac{10i\alpha/9}{1+2i\alpha}\right)\right]
\label{anon}
\end{equation}
where $k=\sqrt{(k_z^2+k_\perp^2)}$ is the total wavenumber, in which $k_z$ and $k_\perp$ are the wavenumbers along and transverse to the confining magnetic field, while $\alpha$ is related to the collision frequency $\nu$ in Eq. (\ref{dougnu}) evaluated at the initial time $t=0$, i. e. with density $n$ and temperature $T$ evaluated at the initial equilibrium; at $t=0$ density and temperature are homogeneous in space and have values $n_0=n(t=0)=1$ and $T_0=T(t=0)=1$ in scaled units, thus at $t=0$ $\nu$ corresponds to  $\nu_0$ and, consequently, $\alpha=\nu_0 k/k_z$.

For weak collisionality ($\alpha \ll 1$), Eq. (\ref{anon}) reduces to:
\begin{equation}
\Re(\omega)\simeq \frac{k_z}{k}\left[1+\frac{3}{2}k^2\right];\;\;\Im(\omega)=-\frac{4}{3}\nu_0k^2
\label{lowalph}
\end{equation}

Moreover, in the case of an infinite plasma (not radially confined, i. e. $k_\perp\to 0$) the above expressions can be easily re-written as:

\begin{equation}
 \Re{(\omega)}=1+\frac{3}{2}k_z^2;\;\;\;\Im{(\omega)}=-\frac{4}{3}\nu_0k_z^2
\label{infplasma}
\end{equation}

We used our Eulerian collisional code (employing the DG operator) to reproduce numerically the collisional damping of plasma waves and to compare the
numerical results to the predictions in Eqs. (\ref{infplasma}). For this numerical experiment we imposed a sinusoidal electric perturbation of amplitude
$E_1=10^{-5}$ and wavenumber $k_1=k_z=0.15$ ($k_1=2\pi/L$ being the fundamental wavenumber) on an initially Maxwellian and homogeneous plasma of electrons (with
fixed ions). The collision frequency $\nu_0$ has been assigned the value $\nu_0=5\times 10^{-4}$. With these choices, the ratio between the Landau damping rate and
the collisional damping rate is $\gamma_{k_1}/\Im(\omega)\simeq 2.5\times 10^{-3}$, that is the effect of Landau damping is negligible when compared to the
collisional dissipation. The results of this simulation are summarized in Fig. \ref{fig7}, in which we reported in semi-log scale the time evolution of the electric
field spectral component $E_{k_{_1}}$. The red-dashed line in the figure represents the theoretical prediction for collisional wave damping in Eq. (\ref{infplasma}):
the numerical results well reproduce the analytical expectation for the collisional damping rate with a relative error of about $2\%$.
\begin{figure}
\epsfxsize=8cm \centerline{\epsffile{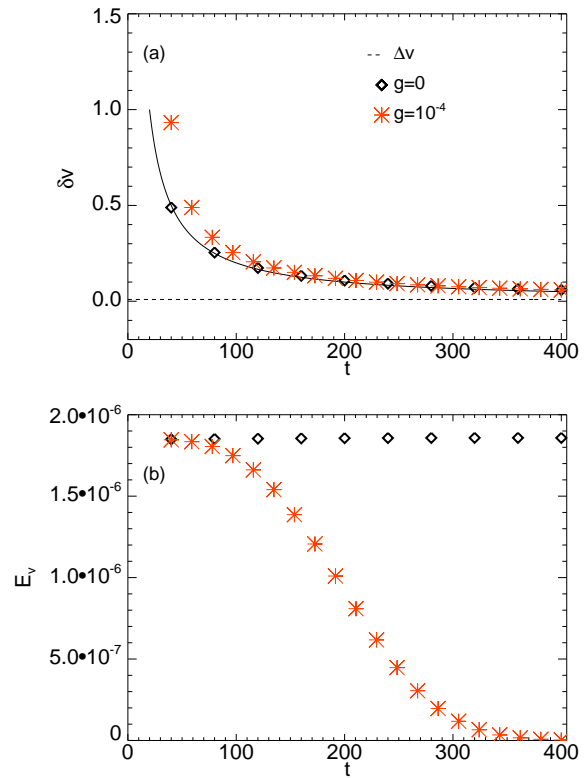}}   % FIGURE N.10
\caption{(Color online) (a) Time evolution of the filamentation scale $\delta v$. The black diamonds refer to the results of a collisionless simulation, while the
red starts represent a collisional case with $g=10^{-4}$. The black solid line is the theoretical curve $1/k_1t$. (b) The spectral energy content $E_v$ of the scale
$\delta v$ as a function of time for the collisionless (black diamonds) and the collisional (red stars) simulations.}
\label{fig10}
\end{figure}

After this preliminary test, in order to point out the effects of the nonlinearities introduced by the DG operator, we repeated the collisional simulations of
plasma wave echoes described in Sec. \ref{echo}, with the parameters $k_1=0.4$, $k_2=0.8$ $\tau=500$, $E_1=E_2=10^{-4}$, and compared the collisional effects of the
DG operator on the echo amplitude to the predictions by O'Neil in Eq. (\ref{d2amp}). The results of this first set of simulations are reported as red stars in Fig.
\ref{fig8} (log-log plot), where the ratio $A/A_{_{NC}}$ is plotted as a function of $\nu_0$. It is evident from this graph that the numerical DG stars depart from
the prediction (black solid line) in Eq. (\ref{d2amp}), in such a way that the decay of the echo amplitude for increasing $\nu_0$ is slower in the case of the DG
operator than in the case of the ON operator in Eq. (\ref{ONeq}).

An analogous effect of reduced collisionality of the DG operator can be also pointed out in the physical process of collisional damping of BGK modes. In Sec.
\ref{ZKBGK}, we compared numerical results and theoretical predictions by Zakharov and Karpman \cite{ZK63} for the collisional dissipation of the electric
oscillations associated to a BGK structure; then, we repeated the same simulation described in Sec. \ref{ZKBGK}, using the nonlinear DG operator to model collisions.
The results of this new simulation are displayed in the semi-log plot of Fig. \ref{fig9} where we reported the time evolution of the electric field spectral
component $E_{k_1}$. As in the simulation described in Sec. \ref{ZKBGK}, the BGK mode is excited using the external driver electric field in Eq. (\ref{drivBGK}) that
is turned off at $t\simeq 300$. Collisions are turned on at $t>500$ (vertical blue-dashed line in the figure). At $t=500$, the value of the collision frequency $\nu$
in Eq. (\ref{dougnu}) has been set $\nu=2\times 10^{-5}$, which corresponds to the value of $\nu$ used in Section \ref{ZKBGK} for the ZK collisional operator. In Fig.
\ref{fig9}, the blue-solid line represents the decay of the electric field amplitude when the ZK operator is used to model collisions, the black-dashed line is the
analytical prediction for collisional damping in Eq. (\ref{gceq}), while the red-solid line indicates the results for the electric field amplitude from the new
simulation in which collisions are modeled through the DG operator. Also here, it is evident that the DG operator produces collisional effects whose strength is
reduced with respect to the case in which the ZK operator is considered.

A possible explanation for this reduced collisionality of the DG operator, both observed in echo and BGK simulations, can be argued through simple physical
arguments: the new ingredient introduced by the DG operator with respect to ON and ZK operators consists in the nonlinearity. In fact, letting $\nu$ be dependent on
$f$ [see Eq. (\ref{dougnu})] implies that the collision frequency varies as the distribution function is shaped by collisions and by other kinetic processes. In
particular, when collisions are at work, the amplitude of the density fluctuations is dissipated in time, while the plasma temperature increases; this produces an
effective collision frequency whose value decreases in time, thus making the collisional damping of plasma waves less efficient with respect to the case (ON or ZK
operator) in which $\nu=const$.

\section{The filamentation problem in the Eulerian algorithms}
As discussed in Sec. \ref{echo}, when particles interact resonantly with a plasma wave, the wave amplitude can be damped in time \cite{landau46}. In the absence of
collisions, once the wave has been Landau damped away, a ballistic term $\propto e^{(-ikvt)}$ is generated in the perturbed distribution function and never
disappears, since each particle, perturbed at $t=0$, carries the memory of its perturbation for all time. For increasing time, this ballistic term produces smaller
and smaller velocity scales of order $\delta v\sim 1/kt$. This physical phenomenon is called velocity space filamentation. From the numerical point of view, as
$\delta v$ gets of the order of the mesh size $\Delta v$ in the velocity domain, a Eulerian code, that follows the evolution of the distribution function in phase
space, introduces a fake numerical dissipation that makes the entropy of the system grow unphysically \cite{califano06,carbone07}.

On the other hand, collisions can erase the memory of the initial perturbation of the particles and consequently arrest the production of small velocity scales.
This suggests that, when the effects of collisionality are taken into account, a careful parameter setting could in principle prevent a collisional Eulerian
simulation to suffer from filamentation. In order to show this, we numerically analyzed the generation of small velocity scales in a Eulerian simulation of linear
Landau damping, both in the absence and in the presence of particle collisions. In this simulation, we imposed a sinusoidal electric perturbation of amplitude
$E_1=10^{-4}$ and wavenumber $k_1=\pi/10$ on an initially Maxwellian and homogeneous plasma of electrons (with fixed ions). At each time step in the simulation we
evaluated $\delta f(x,v,t)=f(x,v,t)-f_{_{M}}(x,v,t)$, where $f_{_M}$ is the associated equivalent Maxwellian distribution computed from the velocity moments of $f$
(density, mean velocity and temperature). The perturbed distribution $\delta f$ has been then Fourier transformed in $v$ to get $\delta \hat{f} (x,\chi,t)$. We have then computed the value of $\chi_m(x,t)$, corresponding to the maximum value of $|\delta f(x,\chi,t)|^2$, and for each time we have determined the velocity associated to the smallest structures as $\delta v(t)=\min_x{\{\chi_m(x,t)\}}$.
\begin{figure}
\epsfxsize=8cm \centerline{\epsffile{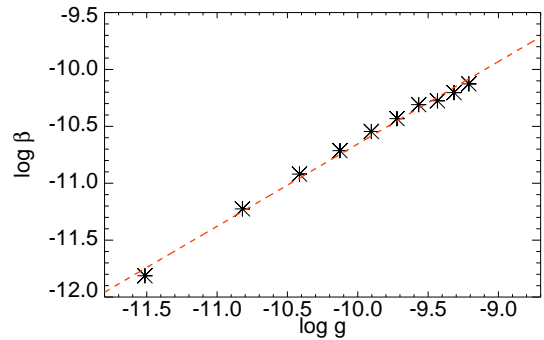}}   % FIGURE N.11
\caption{(Color online) $\log \beta$ versus $\log g$. The black stars indicate the numerical results, while the red-dashed line is the best-fit curve to the numerical starts.}
\label{fig11}
\end{figure}

The numerical results for $\delta v$ versus time are reported in Fig. \ref{fig10}(a). Here, the black diamonds corresponds to the results of a collisionless
simulation, while the red stars to a simulation in which the DG collisional operator has been employed with a value for the plasma parameter $g=10^{-4}$
(corresponding to $\nu_0\simeq 1.2\times 10^{-5}$); moreover, the black solid line represents the curve $1/k_1t$ and the black-dashed line marks the value of the
velocity mesh size $\Delta v$. As one can realize from this picture, in the absence of collisions the black diamonds closely follow the analytical expectation for
the time evolution of the filamentation velocity scales; the red stars for the collisional simulation depart a little from the theoretical curve at early times, but
superpose to the red stars for large times. Finally, at $t=400$ in Fig. \ref{fig10}(a), the filamentation velocity scale $\delta v$ has almost reached the value of
the mesh size $\Delta v$, both in the collisionless and in the collisional case.

However, from the analysis of the numerical results we realized that collisional effects work to dissipate the filamentation scales. This can be appreciated in panel
(b) of Fig. \ref{fig10} in which we plotted the spectral energy content $E_v$ of the scale $\delta v$ as a function of time for the collisionless (black diamonds)
and the collisional (red stars) simulations. While the energy associated with the filamentation scale remains constant in time in the absence of collisions, the
collisional simulation displays an exponential time decay of this energy. At this point, one can fit this exponential decay curve to the function $y=e^{-\beta t^2}$
to get the value of the decay parameter $\beta$. Moreover, by using the fact that $\delta v=1/k_1t$ one gets the relationship between the filamentation scale and the
wavenumber through the parameter $\beta$ as $\delta v=\beta^{1/2}/k_1$. By performing several collisional simulations with different values of the plasma parameter
$g$, we realized that the decay parameter $\beta$ is related to $g$ according to the equation $\beta=Cg^m$, as it can be appreciated in Fig. \ref{fig11} in which we
plotted $\log\beta$ as a function of $\log g$.
Through a best fit procedure on the numerical data of Fig. \ref{fig11} (the best fit curve is represented by the red-dashed line), we evaluated the constants $C$ and
$m$, used to determine the dependence of $\delta v$ on $g$ as $\delta v=0.1811g^{0.362}/k_1$. Obviously, one can use the previous equation to choose the set of
numerical parameters in such a way to prevent the filamentation problem in a collisional Eulerian simulation: once fixed the plasma collisionality $g$ and the value
of the wavenumber $k_1$, the minimum number of gridpoints necessary to discretize velocity space must be such that $\Delta v < 0.1811g^{0.362}/k_1$.
\\
\section{Summary and conclusions}
Understanding the problem of plasma heating both in natural environments (like the solar wind) and in laboratory systems requires to design theoretical models which include the effects of plasma collisionality. The description of collisions in a plasma is a hard goal to achieve both from the analytical and the numerical point of view, mainly due to the nonlinear nature of the Landau collisional operator and to its multi-dimensionality. Numerical algorithms \cite{filbet02,crouseilles04} designed to attach the full Landau problem in 3D velocity space are inexorably limited in resolution and have the problem to introduce unphysical binary collisions.

In this paper we presented a 1D-1V Eulerian collisional Vlasov-Poisson code suitable for the analysis of the propagation of electrostatic waves in collisional plasmas. In this model, plasma collisionality is taken into account by including in the right-hand side of the Vlasov equation 1D differential operators of the Fokker-Planck type, both linear and nonlinear. The numerical algorithm is based on the time splitting method proposed in Ref. \cite{filbet02}. The reduced dimensionality and the simplified form of the differential collisional operators allow to run simulations with high resolution in the phase space numerical domain, thus giving the possibility of describing numerically the effects of collisions on the propagation of electrostatic plasma waves, with a significant degree of physical realism.

We discussed in great details the accuracy of the collisional Vlasov-Poisson algorithm by comparing systematically the numerical results to the analytical
predictions for the collisional effects in unmagnetized plasmas of electrons with fixed ions. Moreover, we implemented the Dougherty collisional operator and
analyzed the effects that the intrinsic nonlinearity of this operator introduces with respect to linear operators. The Dougherty operator, due to its nonlinear nature and to the fact that it takes into account the feedback of collisions on the plasma through the collision frequency, is the only one, among the three discussed in the present work, which can be successfully employed to provide a quite realistic (although approximated) description of a strongly nonlinear turbulent collisional plasma in 1D-1V phase space.

The Dougherty operator has been recently considered \cite{anderson07} to study the effects of collisions on electron plasma waves in nonneutral plasma columns. For these reasons, we expect that our numerical simulations can provide significant support to the interpretation of experimental results of electrostatic plasma waves, when particle collisional effects cannot be neglected.

\section*{Acknowledgments}
The numerical simulations discussed in this paper were  performed on the FERMI supercomputer at CINECA (Bologna, Italy), within the European project PRACE
Pra04-771. D.P. is supported by the Italian Ministry for University and Research (MIUR) PRIN 2009 funds (grant number 20092YP7EY).

%%%%%%%%%%%%%%%%%%%%%%%%

\end{document}